\documentclass[12pt]{article}

\usepackage{natbib}

\usepackage{amsmath}
\usepackage{amssymb}
\usepackage{graphicx}
\usepackage{authblk}
\usepackage{xcolor}
\usepackage{setspace}
\usepackage[margin = 1in]{geometry}
\usepackage{caption}
\usepackage{subcaption}
\usepackage{booktabs}
\usepackage{siunitx}
\usepackage{dsfont}

\usepackage{url}

\usepackage{lineno}

\newcommand{\bs}{\boldsymbol{s}}

\newcommand{\bx}{\boldsymbol{x}}
\newcommand{\bbeta}{\boldsymbol{\beta}}
\newcommand{\bw}{\boldsymbol{w}}
\newcommand{\by}{\boldsymbol{y}}

\newcommand{\boldeta}{\boldsymbol{\eta}}

\newcommand{\bSigma}{\boldsymbol{\Sigma}}
\newcommand{\bQ}{\boldsymbol{Q}}
\newcommand{\bH}{\boldsymbol{H}}

\newcommand{\bB}{\boldsymbol{B}}
\newcommand{\bm}{\boldsymbol{m}}
\newcommand{\bD}{\boldsymbol{D}}
\newcommand{\bone}{\boldsymbol{1}}

\newcommand{\bomega}{\boldsymbol{\omega}}
\newcommand{\bOmega}{\boldsymbol{\Omega}}

\newcommand{\bp}{\boldsymbol{p}}
\newcommand{\bpsi}{\boldsymbol{\psi}}
\newcommand{\bmu}{\boldsymbol{\mu}}
\newcommand{\bb}{\boldsymbol{b}}
\newcommand{\bW}{\boldsymbol{W}}

\newcommand{\bY}{\boldsymbol{Y}}
\newcommand{\bPsi}{\boldsymbol{\Psi}}
\newcommand{\bP}{\boldsymbol{P}}
\newcommand{\bzero}{\boldsymbol{0}}
\newcommand{\bGamma}{\boldsymbol{\Gamma}}
\newcommand{\bgamma}{\boldsymbol{\gamma}}
\newcommand{\bell}{\boldsymbol{\ell}}
\newcommand{\bZ}{\boldsymbol{Z}}
\newcommand{\bd}{\boldsymbol{d}}

\newcommand{\iid}{\overset{\mathrm{iid}}{\sim}}

\newcommand{\bmat}{\begin{bmatrix}}
\newcommand{\emat}{\end{bmatrix}}

\newcommand{\vecc}{\mathrm{vec}}

\newcommand{\E}{\mathrm{E}}

\newcommand{\diag}{\mathrm{diag}}

\newcommand{\pg}{P\'olya-Gamma~}

\title{A reduced rank model for spatial categorical data with many classes}

\author[1*]{Paul B. May}
\author[2]{Andrew Simpson}
\author[2]{Semhar Michael}

\affil[1*]{South Dakota School of Mines \& Technology\protect\\
Department of Electrical Engineering \& Computer Science\protect\\
Rapid City, South Dakota, United States \protect\\
paul.may@sdsmt.edu \vspace{10pt}}
\affil[2]{South Dakota State University\protect\\
Department of Mathematics \& Statistics \protect\\
Brookings, South Dakota United States}
\date{}

\begin{document}

\maketitle

\begin{abstract}
    We develop an identifiable reduced-rank spatial multinomial model for categorical data with many classes. The model represents class-specific spatial effects through a low-dimensional set of shared latent factors, substantially reducing parameter dimension while preserving joint dependence across classes. Because standard conjugate and Pólya–Gamma methods fail under this factorization, we propose a Gibbs sampler using Laplace-approximation proposals within Metropolis–Hastings updates. Simulation studies examine dimension selection and the accuracy of the Laplace proposals. An application to dominant tree species mapping in the Blue Ridge Mountains demonstrates scalable inference and flexible joint predictions for individual classes, class unions, and area-level summaries.
\end{abstract}

%%%%%%%%%%%%%%%%%%%%%%%%%
%%%%%%%%%%%%%%%%%%%%%%%%%

\section{Introduction}

% \begin{enumerate}
%     \item Spatial categorical data
%     \begin{enumerate}
%         \item Oh it's out there.
%         \item Many-class scenarios increasingly common.
%         \item Challenge: spatial dependence, categorical constraints and high dimensional parameter space
%     \end{enumerate}
%     \item LGM's
%     \begin{enumerate}
%         \item Capture spatial dependence, flexible, interpretable.
%         \item Examples of LGMs for multinomial/categorical in the literature.
%         \item One GP per class
%     \end{enumerate}
%     \item Dimensionality and scaling with number of classes
%     \begin{enumerate}
%         \item Many classes = Many params
%         \item Spatial factor model
%     \end{enumerate}
%     \item Inference
%     \begin{enumerate}
%         \item Polya-Gamma
%         \item Laplace Approximations
%     \end{enumerate}
% \end{enumerate}

Spatial categorical data are prevalent in many domains, such as land cover mapping \citep{anderson1976land, lillesand2015remote}, soil taxonomy \citep{mcbratney2003digital}, and species distribution studies \citep{elith2006novel}. These applications often leverage spatially sparse in situ measurements to predict classifications at unobserved locations. Modeling such data is challenging due to the spatial dependencies, compositional constraints, and frequent need for rigorous uncertainty quantification. 

Latent Gaussian models (LGMs) are popular for many forms of spatial data \citep{rue2005gaussian, rue2009approximate, banerjee2014hierarchical}. Latent Gaussian effects are used to model spatial dependencies, often with an autoregressive structure or a Gaussian process, while the data likelihood conforms to properties of the observed variable. Bayesian inference is commonly used for LGMs and other spatial hierarchical models, as it naturally allows uncertainty in the latent effects to be propagated through to predictions and derived quantities. 

LGMs have been applied to categorical and multinomial data \citep{kazembe2007bayesian, finley2009hierarchical, cao2011multinomial, jin2013spatial}, but often with few potential classes. As the number of potential classes increases, the dimension of the model unknowns can increase rapidly, which is concerning for both statistical and computational efficiency. An LGM could accommodate a unique spatial effect for every additional class. While this grants substantial flexibility in modeling the joint probability surface across space, it requires inference on all the spatial effects and potentially their cross-covariances. Further, such flexibility may not be warranted if the class probabilities exhibit dependencies beyond their natural constraint of summing to unity. The patterns of presence/absence across even many classes can be driven by a few latent factors, e.g., temperature, precipitation, and soil gradients simultaneously favoring some tree species while inhibiting others. This motivates the use of a spatial factor model \citep{christensen2002latent, ren2013hierarchical, taylor2019spatial}, where the multivariate spatial process is dependent on a relatively small set of shared factors, creating a reduced-rank linear predictor. 

A computational challenge to Bayesian inference on LGMs is integrating across the high-dimensional latent effects. \cite{polson2013bayesian} developed a \pg data augmentation scheme for binomial logistic models where the conditional posterior of the latent effects, given auxiliary \pg variables, is Gaussian, enabling an efficient Gibbs sampler. \cite{bradley2019spatio} instead specified a multivariate logit-beta prior for the latent effects (not an LGM in the strict sense, but allowing similar prior spatial structures) which is conjugate to the binomial logistic likelihood. Both techniques can be generalized to multinomial logistic models by exploiting conditional one-versus-the-rest binomial likelihoods. However, this requires a component-wise separability of the linear predictor, where the log-odds of each class depends on a unique latent effect. This precludes a factor model, where the log-odds for each class depends on a shared set of latent effects. The \pg technique can be salvaged in this scenario by using a ``stick-breaking'' link function instead of the multinomial logistic link function \citep{linderman2015dependent}. However, the stick-breaking link is highly asymmetric and dependent on the class ordering, and for most applications there is no natural ordering of the classes. Alternatively, the conditional posterior of the latent effects can be approximated with Laplace approximations. This is most notably employed in the Integrated Nested Laplace Approximation (\citealp{rue2009approximate}; INLA), where a Laplace approximation is substituted for the conditional posterior of the latent effects, integrating across the remaining parameters through a deterministic quadrature. Software package `R-INLA' (\url{www.r-inla.org}) can fit multinomial LGMs with or without shared spatial factors. However, even with a modest number of classes and small number of latent factors, it is difficult to accurately depict the posterior distribution of the factor weights through a deterministic quadrature. 

We propose an identifiable logistic multinomial model with reduced-rank spatial effects. Computationally, inference is conducted through a Gibbs sampler. To sample the latent spatial effects, we use Laplace approximations to generate proposals for a Metropolis-Hastings step. The same procedure is used to sample the factor weights. This inference scheme does not require separability of the linear predictor and avoids deterministic quadratures, remaining tractable for data with many classes. The focus of this work is on categorical data, but the model and computational techniques are applicable to multinomial models with an arbitrary, but known, number of trials.

The rest of the paper is organized as follows. Section \ref{sec:methods} introduces a spatial categorical model, the reduced-rank counterpart, and a Gibbs sampler for posterior inference. Section \ref{sec:sim} contains simulation studies examining dimension selection, the model's predictive performance when a reduced-rank structure is present in the data, and the accuracy of the Laplace proposals. Section \ref{sec:rda} demonstrates the method on a practical dataset, predicting the probability of species dominance across 24 tree species groups within the Blue Ridge Mountains. Section \ref{sec:disc} provides further discussion on the work and possible extensions.

%%%%%%%%%%%%%%%%%%%%%%%%%%%
%%%%%%%%%%%%%%%%%%%%%%%%%
\section{Methods} \label{sec:methods}
%%%%%%%%%%%%%%%%%%%%%%%%%
\subsection{Spatial categorical model}\label{sec:methods:fullrank}

Consider the categorical process model for $J$ classes
\begin{align}
    \by(\bs) &\sim \mathrm{Categorical}\left(\bp(\bs)\right), \label{eq:catproc1}\\ 
    \bp(\bs) &= \mathrm{softmax}_J\left(\bpsi(\bs)\right),\\
    \bpsi(\bs) &= \bmu + \boldeta(\bs), \label{eq:catproc3}
\end{align}
for all locations $\bs$ in some spatial domain $\mathcal{D}$, where 
\begin{equation}
    \mathrm{softmax}_J(\bpsi) = \left[\frac{\exp(\psi_j)}{1 + \sum_{\ell=1}^{J-1}\exp(\psi_{\ell})}\right]_{j =1,\ldots,J-1}
\end{equation}
with $\bpsi(\bs)\in\mathbb{R}^{J-1}$ being a vector of logits, $\bmu\in\mathbb{R}^{J-1}$ is a logit-scale mean vector and $\boldeta(\bs)\in\mathbb{R}^{J-1}$ is a multivariate Gaussian process. Here we have arbitrarily set class $J$ as our control class, with $p_J(\bs)$ uniquely determined by the other $J-1$ probabilities.

An important feature of this model is the joint separability of the logits and effects,
\begin{equation}
    \bpsi_j(\bs)=\mu_j+\eta_j(\bs);\quad\text{for}~~j=1,\ldots,J-1,
\end{equation}
where each logit is a function of a distinct set of effects. Considering posterior inference, this is necessary for the \pg sampler of \cite{polson2013bayesian} and logit-beta conjugate prior of \cite{bradley2019spatio}. Both methods require that, conditional on the remaining effects, the likelihood for $(\mu_j, \eta_j(s))$ is binomial and a well-defined conditional prior exists for these components.

Gaussian process $\boldeta(\bs)$ is intended to capture spatial patterns in the data, reflecting the prior belief that locations closer together have more similar class probabilities. For the rest of our development, we will assume a fixed-rank Gaussian process,
\begin{align}
    \boldeta(\bs) &= \bZ^T \bb(\bs),\\
    \vecc(\bZ) &\sim \mathrm{MVN(\bzero, \bSigma \otimes \bQ^{-1})},
\end{align}
where $\bb(\bs)\in \mathbb{R}^k$ is a vector of spatial basis functions and $\bZ\in\mathbb{R}^{k\times(J-1)}$ is a matrix of spatial weights with between-class covariance $\bSigma$ and spatial precision $\bQ$. The spatial weights represent random effects at a discrete set of $k$ locations, while the basis functions project these effects onto the continuous spatial domain. In particular, we use the predictive process of \cite{banerjee2008gaussian}: First, $k$ spatial knots, $\bell_1,\ldots,\bell_k\in\mathcal{D}$, are fixed across the study domain. The spatial precision matrix and basis functions are generated by any positive-definite kernel; we used the exponential correlation function, 
\begin{align}
    Q_{ij} &= \exp\left(-\frac{\|\bell_i - \bell_j\|}{\phi}\right)\quad\text{for}~~ i,j=1,\ldots,k,\\
    b_i(\bs) &= \exp\left(-\frac{\|\bs - \bell_i\|}{\phi}\right)\quad\text{for}~~ i = 1,\ldots,k.
\end{align}
Parameter $\phi$ is a range parameter controlling the rate of decay of spatial correlation with distance. Both the spatial precision matrix and basis functions depend on this unknown parameter, but we temporarily omit this dependence from our notation. Many other fixed-rank processes are possible, such as Wendland basis functions \citep{nychka2015multiresolution} and the stochastic partial differential equation method \citep{lindgren2011explicit}, and the choice of process does not affect the main developments in this work. Fixing the rank of the spatial process is a computational convenience, and our primary focus is reducing the rank of between-class relationships.

%%%%%%%%%%%%%%%%%%%%%%%%%

\subsection{Reduced-rank spatial categorical model}

For the categorical model in equations (\ref{eq:catproc1}--\ref{eq:catproc3}), the dimension of the unknown parameters and effects increases considerably with the number of classes. While the spatial precision $\bQ$ typically depends on few parameters (e.g., just a spatial range parameter) and can provide strong regularization across the rows of $\bZ$, we still need to depict the variation of $J-1$ marginal Gaussian processes and the covariance between them via $(J-1) \times (J-1)$ matrix $\bSigma$. 

However, if the data-generating process induces strong dependencies between class probabilities beyond the simplex constraints, $\sum p_j(s)=1$, we can substantially reduce the dimension by factoring the spatial effects,
\begin{equation}
    \bpsi(\bs) = \bmu + \bGamma\tilde{\boldeta}(\bs), \label{eq:redranklogits}
\end{equation}
where $\bGamma \in \mathbb{R}^{(J-1)\times u}$ is an unknown factor matrix with rank $u\leq J-1$ and $\tilde{\boldeta}(\bs)\in\mathbb{R}^u$ is a multivariate Gaussian process. Under this formulation, the latent spatial variation is driven by $\tilde{\boldeta}(\bs)$ while the induced process, $\boldeta(\bs)=\bGamma\tilde{\boldeta}(\bs)$, lies in a $u$-dimensional subspace and is degenerate when $u<J-1$. 

We maintain the fixed-rank process on $\tilde{\boldeta}(\bs)$ with
\begin{align}
    \tilde{\boldeta}(\bs) &=\bW^T\bb(s), \\
    \vecc(\bW)&=\mathrm{MVN(\bzero,~\bOmega^{-1} \otimes \bQ^{-1}}),
\end{align}
where $\bW\in \mathbb{R}^{k\times u}$ is a factored matrix of spatial weights with row-precision $\bOmega$. The primary spatial weights are now $\bW$ while the induced weights $\bZ = \bW\bGamma^T$ are degenerate for $u<J-1$ with rank-$u$ row-covariance $\bSigma=\bGamma\bOmega^{-1}\bGamma^T$.

For unconstrained $\bGamma,\bOmega$, the factorization $\bSigma=\bGamma\bOmega^{-1}\bGamma^T$ is not unique, so some constraints on $\bGamma,\bOmega$ are necessary for identifiability. We assume $\bGamma$ is unit-lower triangular, 
\begin{equation}
    \Gamma_{ij} = 
    \begin{cases}
        1\quad& i=j \\
        0\quad& i <j\\
        \text{free} \quad& \text{otherwise}
    \end{cases}
\end{equation}
and $\bOmega = \diag(\bomega)$ with marginal precisions $\omega_j;\,j=1,\ldots,u$. This factorization is unique and comprehensive, in the sense that there is a bijection, $(\bGamma, \bOmega) \leftrightarrow \bGamma\bOmega^{-1}\bGamma^T$, between the factors and the set of rank-$u$ PSD matrices.

If $u \ll J-1$, the parameter reduction can be substantial. Instead of inferring the $J(J-1)/2$ unconstrained parameters of full-rank $\bSigma$, we infer $(J-1)u - u(u-1)/2$ parameters of $\bGamma$ and $\bOmega$, a reduction of $(J-u-1)(J-u)/2$ parameters. This reduction would be the same for any unique decomposition of rank-$u$ PSD matrices. Instead of inferring the variation of $J-1$ Gaussian processes, we infer the variation of $u$ Gaussian processes. 

This parsimony comes at the cost of strong assumptions. For fixed $\bmu$ and $\bGamma$, the possible probability vectors lie in a $u$-dimensional submanifold of the simplex. This makes the selection of latent dimension $u$ quite important. If $u$ is too small, plausible probability tuples will be impossible, and model predictions may exhibit complex biases. If $u$ is too large, inference and computation will be inefficient. We examine dimension selection through simulation in Section \ref{sec:sim:dim}, finding that if the data-generating process is of reduced rank, selecting $u$ through typical model selection techniques is preferable to defaulting to a full-rank model.

If $u < J-1$, the posterior sampling methods of \cite{polson2013bayesian} and \cite{bradley2019spatio} are no longer applicable. Each component process $\tilde{\eta}_j(\bs)$ potentially affects all logits, and the induced processes $\eta_j(\bs)$ have no well-defined conditional prior. We instead use Laplace approximations as proposals to Metropolis-Hastings steps within a Gibbs sampler. 
%%%%%%%%%%%%%%%%%%%%%%%%%
\subsection{Posterior inference}

For the reduced-rank model in (\ref{eq:redranklogits}), the unknown parameters/effects are mean vector $\bmu$, spatial effects $\bW=[\bw_1\cdots\bw_u]$, marginal precisions $\bomega$, the unconstrained entries of the factor matrix $\bgamma \leftrightarrow\bGamma$, and the range parameter $\phi$, determining the spatial precision matrix and basis functions, $\bQ(\phi),~\bb(\bs|\phi)$. 

Given data observations $\bY = [\by(\bs_1)\cdots\by(\bs_n)]^T$ and assuming independent priors for $\omega_j;\,j=1,\ldots,u$, the posterior distribution of the unknowns is
\begin{equation}
    f(\bmu,\bW, \bomega, \bgamma,\phi|\bY) \propto f(\bY | \bmu,\bW,\bgamma,\phi)\cdot \pi(\bmu)\cdot\prod_{j=1}^u\left\{\pi(\bw_j|\omega_j,\phi)\cdot\pi(\omega_j) \right\} \cdot\pi(\bgamma)\cdot\pi(\phi). \label{eq:posterior}
\end{equation}
Parameters $\bmu$ and $\bgamma$ are assigned multivariate normal priors with user-defined mean and precision. The spatial effects, $\bw_j|\bomega_j,\phi\,;\,j=1,\ldots u$, have multivariate normal prior with zero mean and precision $\bomega_j\bQ(\phi)$. Precision parameters, $\omega_j\,;\,j=1,\ldots,u$, and range parameter $\phi$ are strictly positive, so independent Gamma or log-normal priors, for example, are appropriate.

We use a Gibbs sampler to draw random samples from (\ref{eq:posterior}). The primary challenge is drawing conditional samples from $\bgamma|\cdots$ and $\bw_j|\cdots\,;\,j=1,\ldots,u$; these vectors can have high dimension and complex distributions. We use Laplace approximations to generate proposals for a Metropolis-Hastings step. Focusing on some $\bw_j$ for exposition, we compute the mode of the log-conditional posterior, $\hat{\bw}_j$, through a Newton-Raphson routine. A proposed sample is then generated through the Laplace approximation, $\bw_j^*\sim \mathrm{MVN}\left(\hat{\bw}_j,\,-\bH(\hat{\bw}_j)^{-1}\right)$, where $\bH(\hat{\bw}_j)$ is the Hessian at the mode. This sample is either accepted or rejected via a Metropolis-Hastings step.

The efficiency of the sampler is primarily determined by the acceptance rate, which is in turn determined by the accuracy of the Laplace approximation. In our experience, the spatial effects $\bw_j|\cdots$ are typically guilty of the lowest acceptance rates, not the factor entries, $\bgamma$. Even for $\bw_j|\cdots$ of high-dimension, acceptance rates can remain high enough for a feasible sampler, unless the unobserved prior precision, $\omega_j$, is very small: If $\omega_j$ is small, $\bw_j$ will have large variance, producing logits of large magnitude. Large logits push the probabilities against the asymptotes of the softmax link function, producing skewed posterior distributions for $\bw_j$ and a posterior expected value further from zero than the posterior mode, e.g., $|\E[\bw_j|\cdots]| \gg |\hat{\bw}_j|$. We demonstrate this effect through simulation in Section \ref{sec:sim:acc}.

A \pg sampler could be used to iteratively sample $\mu_j|\cdots\,;\,j=1,\ldots,J-1$, since $\mu_j$ affects only $\psi_j(\bs)$. However, we chose to again use a Laplace to Metropolis-Hastings procedure to sample $\bmu|\cdots$ simultaneously, as the acceptance rates were near unity in all our analyses and it avoids the additional Gibbs autocorrelation from cycling through $\mu_j$ individually.

Finally, using independent Gamma priors for $\omega_j\,;\,j=1,\ldots,u$, the conditional posteriors for $\omega_j$ are also Gamma distributed, so samples can be drawn exactly. We also declared a Gamma prior for range $\phi$, but here this prior is not conjugate, so we used another Metropolis-Hasting step to sample $\phi|\cdots$.

Full details on the Gibbs sampler can be found in Appendix \ref{app:gibbs}, including the gradients and Hessians with respect to $\bw_j$, $\bgamma$, and $\bmu$ required for Newton-Raphson and the Laplace approximation.

To predict class presence and probabilities at an unobserved location, $\bs^*$, we generate samples from the posterior predictive distribution, $ f\left(\by(\bs_*) | \bY\right)$, using $M$ samples of the model unknowns:
\begin{align}
    \by_{(m)}(\bs^*) &\sim \mathrm{Categorical}(\bp_{(m)}(\bs^*)), \label{eq:pps}\\
    \bp_{(m)}(\bs^*) &= \mathrm{softmax}(\bpsi_{(m)}(\bs^*)), \\
    \bpsi_{(m)}(\bs^*) &= \bmu_{(m)} + \bGamma_{(m)}\bW_{(m)}^T\bb(\bs^*|\phi_{(m)})\quad\text{for}~~m=1,\ldots M.
\end{align}
Often, the analyst is interested in area summaries, such as the probability of any occurrence of class $j$ within a given area. Predictive distributions for such summaries can be approximated by producing observation-level predictive samples (\ref{eq:pps}) on a dense grid across the target area and summarizing these samples accordingly. For example, to infer the probability of any class-$j$ occurrence within area $\mathcal{A}\subset\mathcal{D}$, using grid locations $\bs_1^*,\ldots,\bs_g^*\in\mathcal{A}$, the relevant predictive samples are
\begin{equation}
    y_{j,(m)}(\mathcal{A}) = \mathds{1}\left(\sum_{i=1}^g \left\{y_{j,(m)}(\bs_i^*)\right\}>0 \right)\quad\text{for}~~m=1,\ldots M.
\end{equation}
Further, a primary advantage of a multinomial model over $J$ separate binomial models is the ability to make joint inference across classes. For example, let $\bar{y}_\mathcal{C}(\bs)$ be the binomial variable of total class occurrences within class subset $\mathcal{C}$. The relevant predictive samples are
\begin{equation}
    \bar{y}_{\mathcal{C},(m)}(\bs^*)=\sum_{j\in\mathcal{C} }y_{j,(m)}(\bs^*)\quad\text{for}~~m=1,\ldots M.
\end{equation}
%%%%%%%%%%%%%%%%%%%%%%%%%%
\subsection{Fixed covariate effects}

While not studied in this work, it is possible to include covariate effects in the model,
\begin{equation}
    \bpsi(\bs) = \bmu + \bbeta^T\bx(\bs)+\boldeta(\bs),
\end{equation}
where $\bbeta\in\mathbb{R}^{p\times J-1}$ is a matrix of regression coefficients and $\bx(\bs)\in\mathbb{R}^p$ is a location-specific covariate vector, fixed and observed across the study domain. For many classes and/or covariates, the dimension of $\bbeta$ will be large and some dimension reduction may be desired. The simplest option is a shared factorization,
\begin{equation}
    \bpsi(\bs) = \bmu + \bGamma\left(\tilde{\bbeta}^T\bx(\bs)+\tilde{\boldeta}(\bs)\right),
\end{equation}
where $\tilde{\bbeta} \in \mathbb{R}^{p\times u}$ is a reduced matrix of regression coefficients. This assumes the covariate and spatial effects reside in the same $u$-dimensional subspace, which is restrictive, but introduces few additional unknowns and improves the posterior precision of $\bGamma$. Alternatively, a separate factorization could be assumed,
\begin{equation}
    \bpsi(\bs) = \bmu + \bGamma_x\tilde{\bbeta}^T\bx(\bs)+\bGamma\tilde{\boldeta}(\bs),
\end{equation}
with reduced dimension $u_x$ for the covariate effects. This model is more flexible, but infers an additional factor matrix and potentially requires a two-dimensional model selection among candidate dimensions $(u_x, u)$.
%%%%%%%%%%%%%%%%%%%%%%%%%%
%%%%%%%%%%%%%%%%%%%%%%%%%
\section{Simulation study}\label{sec:sim}
Here we present a simulation study demonstrating the selection of latent dimension $u$ and the effect of the marginal precisions $\omega_j$ on the accuracy of the Laplace proposal. For all simulations, we generated categorical data according to the reduced-rank model (\ref{eq:redranklogits}) with $J=5$ classes and latent dimension $u_\text{true}=2$. The number of potential classes was set to a modest value here to make repeated simulation and exhaustive dimension selection easier; we demonstrate feasibility of the model for many classes in the real data analysis, Section \ref{sec:rda}. Full datasets were generated on a $50 \times 50$ regular grid of locations in $[0,1]\times[0,1]$. A subset of $n=250$ locations was randomly selected as training observations. For the fixed-rank Gaussian processes, spatial knots were placed on a $15\times15$ regular grid. The spatial range was fixed at $\phi=0.2$ for all simulations so that the data consistently exhibited strong spatial patterns. 
%%%%%%%%%%%%%%%%%%%%%%%%%
\subsection{Dimension selection}\label{sec:sim:dim}
\begin{figure}[p]
    \centering
    \begin{subfigure}[b]{\textwidth}
        \centering
        \includegraphics[width=\linewidth]{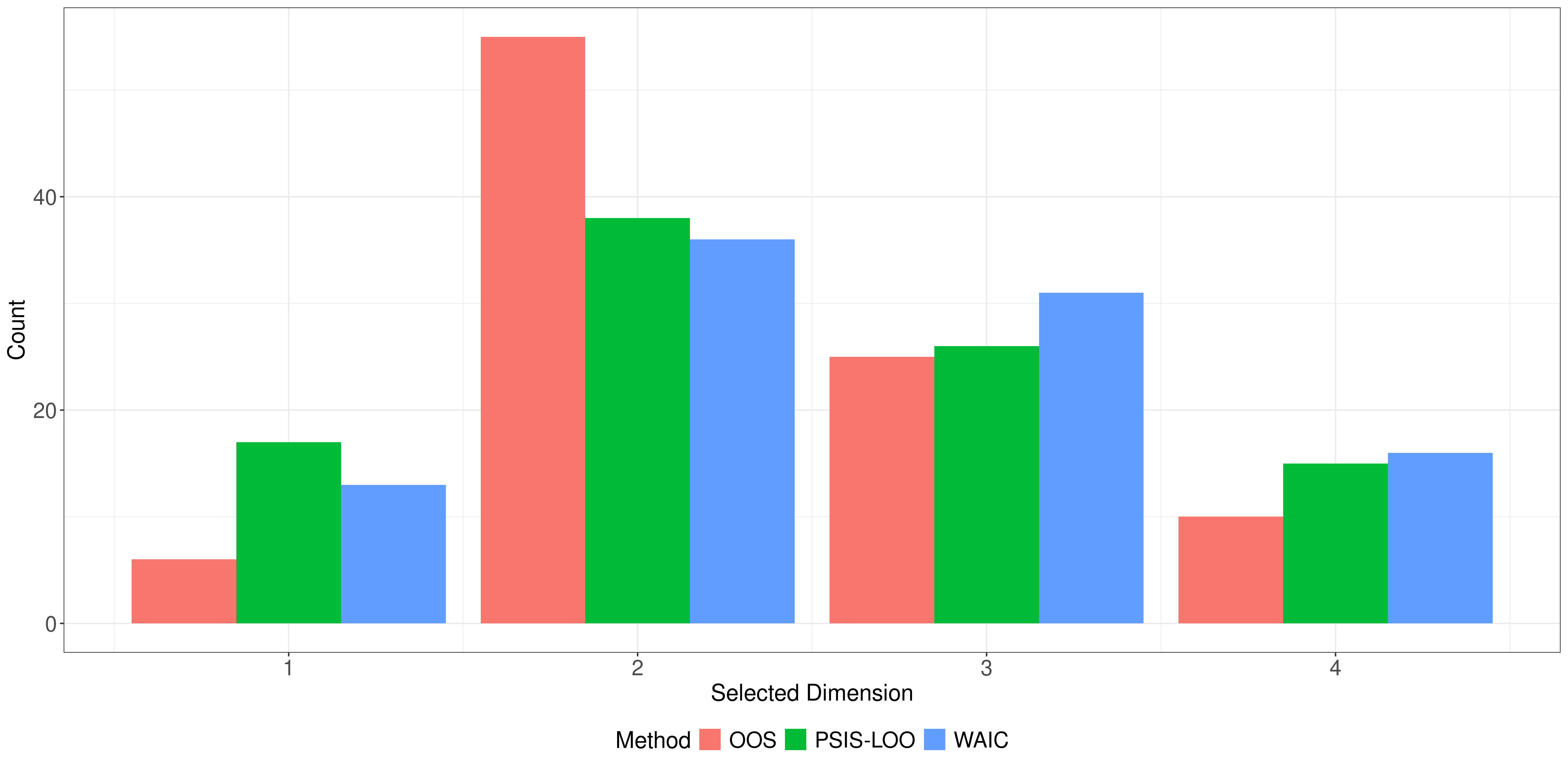}
        \caption{The selected dimension of the latent effects via WAIC, PSIS-LOO, and best out-of-sample (OOS) log-predictive density.}
    \end{subfigure}\\
        \begin{subfigure}[b]{\textwidth}
        \centering
        \includegraphics[width=\linewidth]{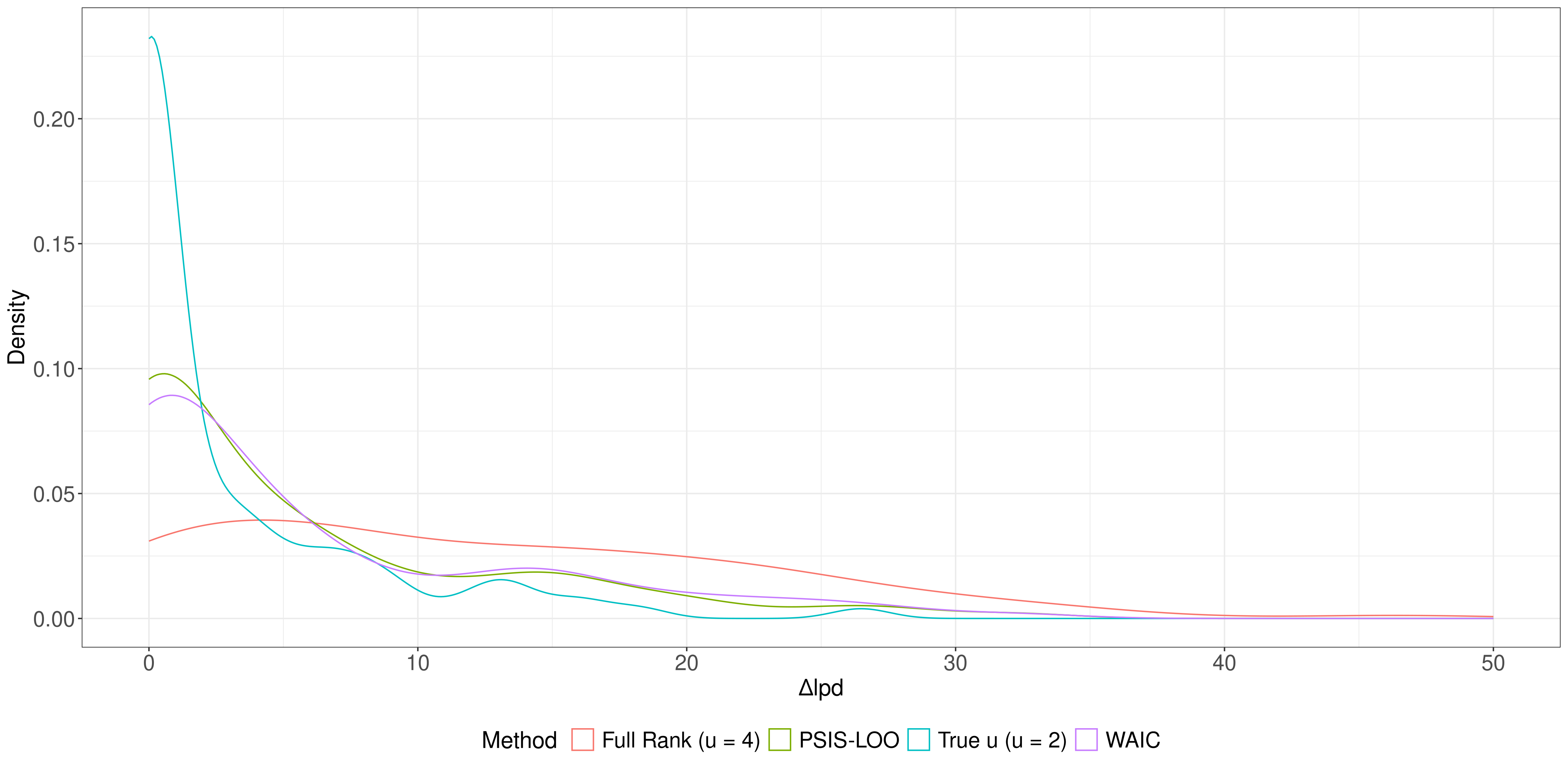}
        \caption{Kernel density estimates of $\Delta\mathrm{lpd}$, the difference in the out-of-sample log predictive density for the best dimension (for that particular realization) and selected dimension. The full-rank model had nine instances of $\Delta\mathrm{lpd}>50$, but the horizontal axis was truncated at 50 for visualization.}
    \end{subfigure}
    \caption{The results of 100 simulations with $J=5$ and $u_\text{true}=2$. The true dimension of the latent effects is often not selected by WAIC and PSIS-LOO, but neither is the true dimension always superior for interpolating a single multivariate realization given limited observations. On average, using the true dimension delivers better predictive performance, but selecting a dimension via WAIC or PSIS-LOO is substantially better than defaulting to a full rank model.}
    \label{fig:dimselect}
\end{figure}

We studied the selection of latent dimension $u$ and the effect on out-of-sample prediction performance. Using the simulation parameters described above, we simulated $t=1,\ldots100$ datasets with $\mu_{j,t}\iid \mathrm{N}(0,1)$, $\gamma_{j,t}\iid \mathrm{N}(0,1)$, and $\omega_{j,t} \iid \mathrm{Gamma}(4,4)$ (shape/rate parametrization). Our priors were chosen to match the above data-generation distributions, except for the fixed spatial range $\phi=0.2$, which was assigned the prior $\phi\sim\mathrm{Gamma}(4,20)$. For each generated dataset, we fit models for $u=1,2,3,4$, drawing 10,000 posterior samples after a burn-in of 3,000 samples. Models were selected by WAIC and PSIS-LOO \citep{watanabe2010asymptotic, vehtari2017practical}, both being estimates of cross-validation log-predictive density. All potential models were tested via the total log-predictive density on the 2,250 test locations:
\begin{equation}
    \mathrm{lpd}(u) = \sum_{i=1}^{2,250}\log(p_u(\by_i^*|\bY)).
\end{equation}
WAIC and PSIS-LOO often did not select the true dimension, $u_\text{true}=2$ (Figure \ref{fig:dimselect}). However, the true dimension did not always yield the best out-of-sample predictive performance. Therefore, we compared the selected models by the difference between the $\mathrm{lpd}$ of the selected model and that of the best model, $\Delta\mathrm{lpd}(u) = \mathrm{lpd}_\text{best}-\mathrm{lpd}(u)$. WAIC yielded a mean $\Delta\mathrm{lpd}$ of $6.04$, with the sample standard deviation of the mean being $0.78$. PSIS-LOO performed similarly, with a mean of $5.98$ and standard deviation of $0.88$. The full-rank model, $u=4$, had a mean of $22.32$ and a standard deviation of $3.81$. The true model, $u = u_\text{true}$, had a mean of $2.89$ and a standard deviation of $0.50$.

\subsection{Accuracy of the Laplace approximation}\label{sec:sim:acc}
\begin{figure}[p]
    \centering
    \includegraphics[width=\linewidth]{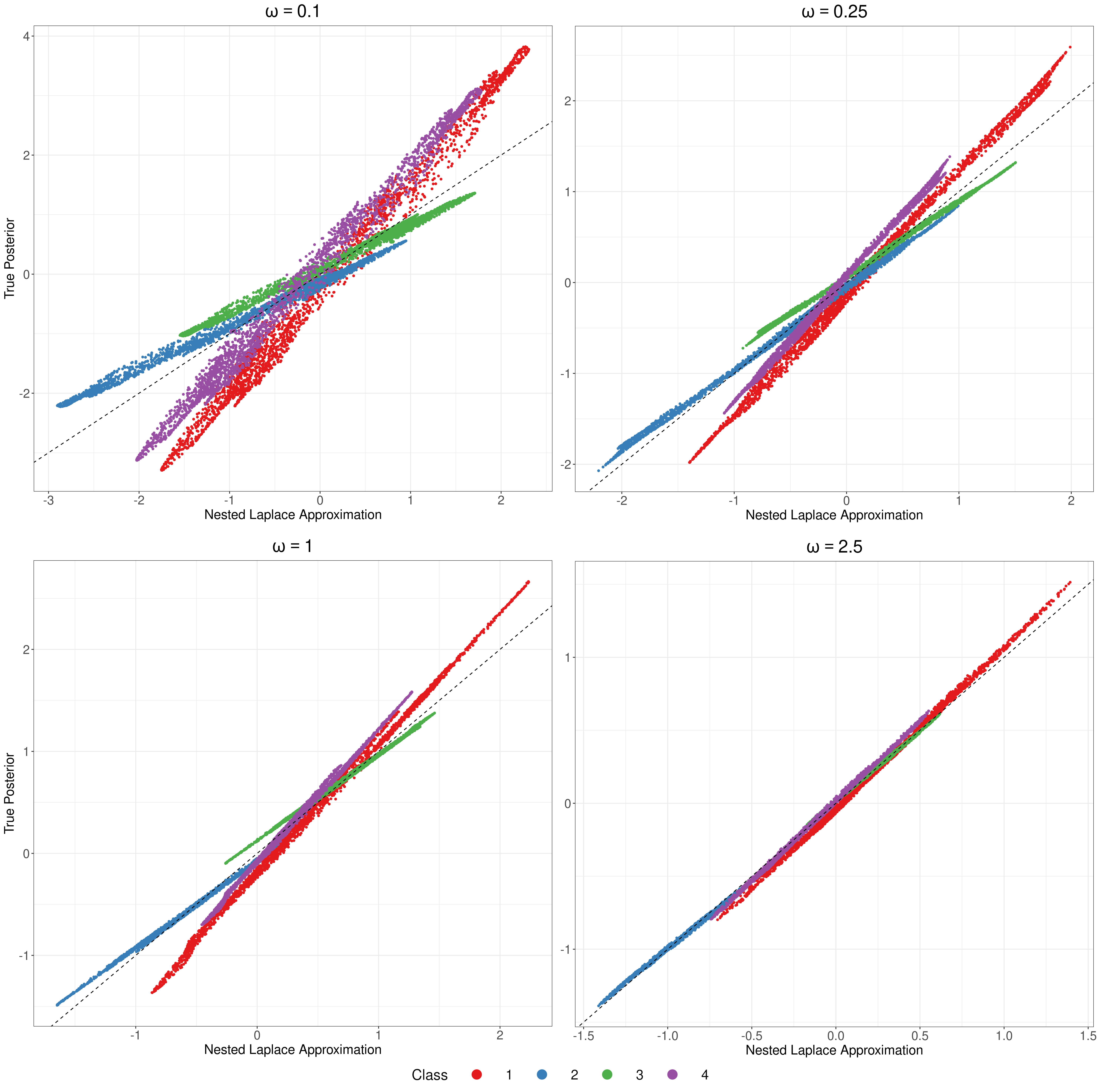}
    \caption{Posterior expected values of the logits with varying marginal precisions, $\omega$, compared between two inference methods. The first method accepts or rejects Laplace approximation proposals of $\bw_j|\cdots;j=1,\ldots,u$ through a Metropolis Hastings step, asymptotically sampling from the true posterior. The second method is a nested Laplace approximation, where the Laplace approximation is always accepted. Lower marginal precisions produce a more severe linear bias in the predictions of the nested Laplace approximation and lower acceptance rates in the Metropolis-Hastings step.}
    \label{fig:omegabias}
\end{figure}

The efficiency of the proposed sampler depends on the acceptance rates for $\bw_j|\cdots\,;\,j=1,\ldots,u$. We found the sharpest determinant of Laplace approximation accuracy and downstream acceptance rates to be the precisions, $\omega_j\,;\,j=1,\ldots,u$, not the rank of the Gaussian processes, $k$. To demonstrate, we simulated data as described in the beginning of Sections \ref{sec:sim} and \ref{sec:sim:dim}, but fixed $\bmu$ and $\bgamma$ as a single realization from their data-generation distributions. We then generated four datasets, setting all $\omega_j$ to $0.1,\,0.25,\,1.0,\,2.5$. For each dataset, we produced posterior samples using the proposed sampler and approximate posterior samples where the Laplace proposals $\bw_j^*|\cdots;\,j=1,\ldots,u$ are always accepted outright with no Metropolis-Hastings criterion, constituting a nested Laplace approximation of the true posterior. The latent dimension $u$ was assumed known and fixed. For both the true posterior and nested Laplace approximation, we produced posterior predictive samples for the unobserved logits, $\bpsi(\bs)$. Relative to the true posterior, the nested Laplace approximation exhibits severe linear bias for small $\omega_j$ and this bias attenuates as $\omega_j$ increases (Figure \ref{fig:omegabias}). In parallel, the acceptance rates for $\bw_j|\cdots$ in the proposed sampler of the true posterior were $26,\,60,\,67,\,83\%$ with increasing $\omega_j$, indicating the proposal distribution is more dissimilar to the true conditional posterior for small precisions.

%\begin{figure}[p]
%    \centering
%    \includegraphics[width=\linewidth]{graphics/prob_omega_merged.png}
%    \caption{Caption}
%    \label{fig:placeholder}
%\end{figure}

%%%%%%%%%%%%%%%%%%%%%%%%%%%%%%
\section{Data Analysis}\label{sec:rda}
\begin{figure}[htpb]
    \centering
    \includegraphics[width=\linewidth]{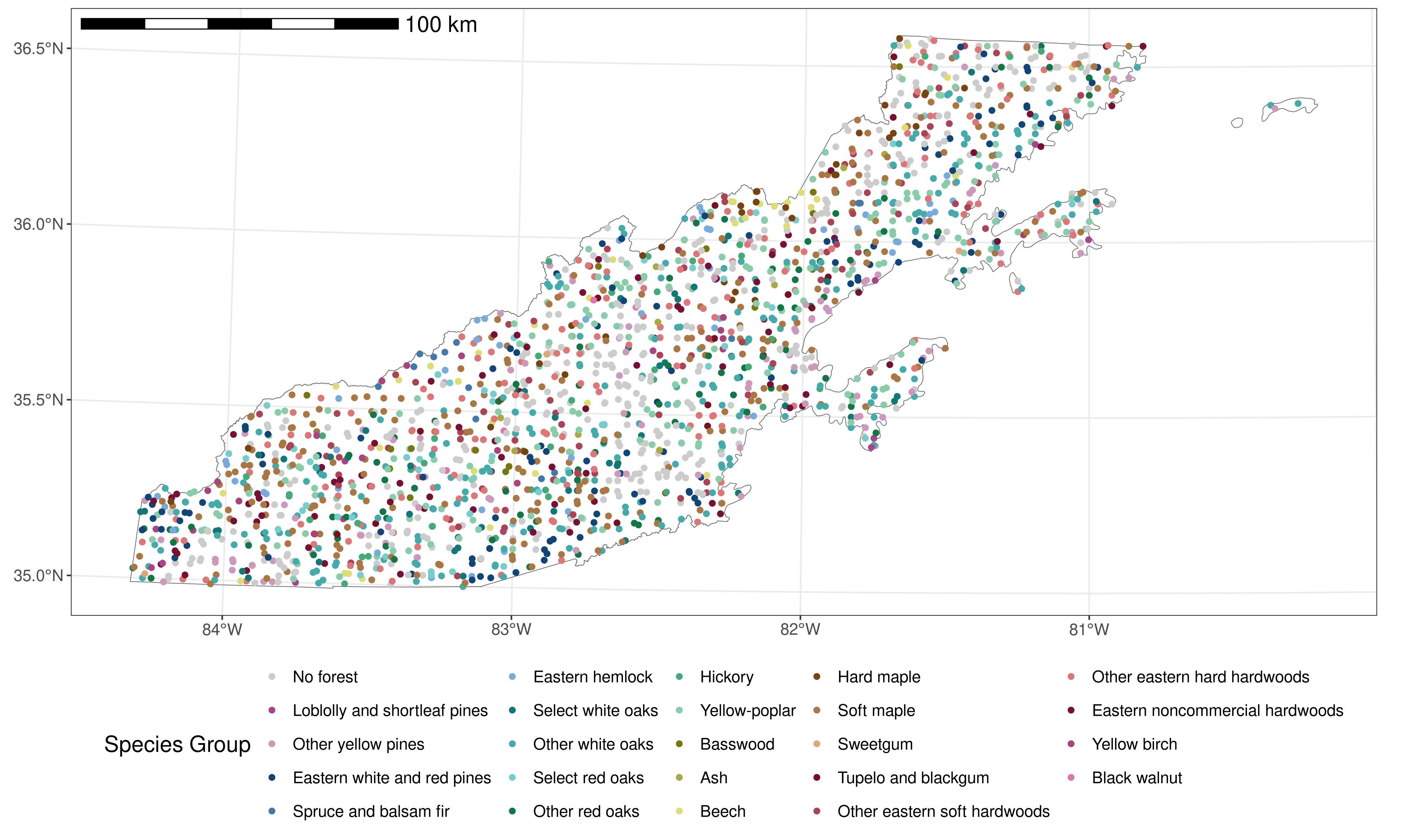}
    \caption{Observations of 24 unique dominant species classes across the Blue Ridge Mountains of North Carolina.}
    \label{fig:traindata}
\end{figure}

We applied the reduced-rank model to predicting dominant tree species across the Blue Ridge Mountains of North Carolina. The Forest Inventory and Analysis (FIA) program of the US Forest Service maintains a permanent network of randomly placed field plots across the contiguous US \citep{bechtold2005enhanced}. The field plots are cyclically revisited and important forest attributes are measured. Using the most recent measurements from the 2,063 plots within our study area, we assigned a dominant species class to each plot through the most frequent FIA major tree species code (Figure \ref{fig:traindata}). Plots with no trees were assigned a `no forest' class. Including the `no forest' class, the study area has 24 unique classifications. Some classes have few observations, with half the classes possessing less than 48 observations and three classes (Ash, Black Walnut, Sweetgum) possessing less than 10 observations. The goal of the analysis is to leverage the spatially sparse field measurements to produce spatially complete predictions of class occurrences.

For the fixed-rank Gaussian processes, $k=1000$ spatial knots were placed by randomly selecting plot locations without replacement. We set the prior distributions $\mu_{j}\iid \mathrm{N}(0,1)$, $\gamma_{j}\iid \mathrm{N}(0,1)$, $\omega_{j} \iid \mathrm{Gamma}(4,4)$, and spatial range $\phi \sim\mathrm{Gamma}(16, 16\cdot10^{-4})$ in meters (corresponding to a prior mean and standard deviation of $\text{10,000}\pm\text{2,500 m}$).

\begin{figure}[htpb]
    \centering
    \includegraphics[width = \linewidth]{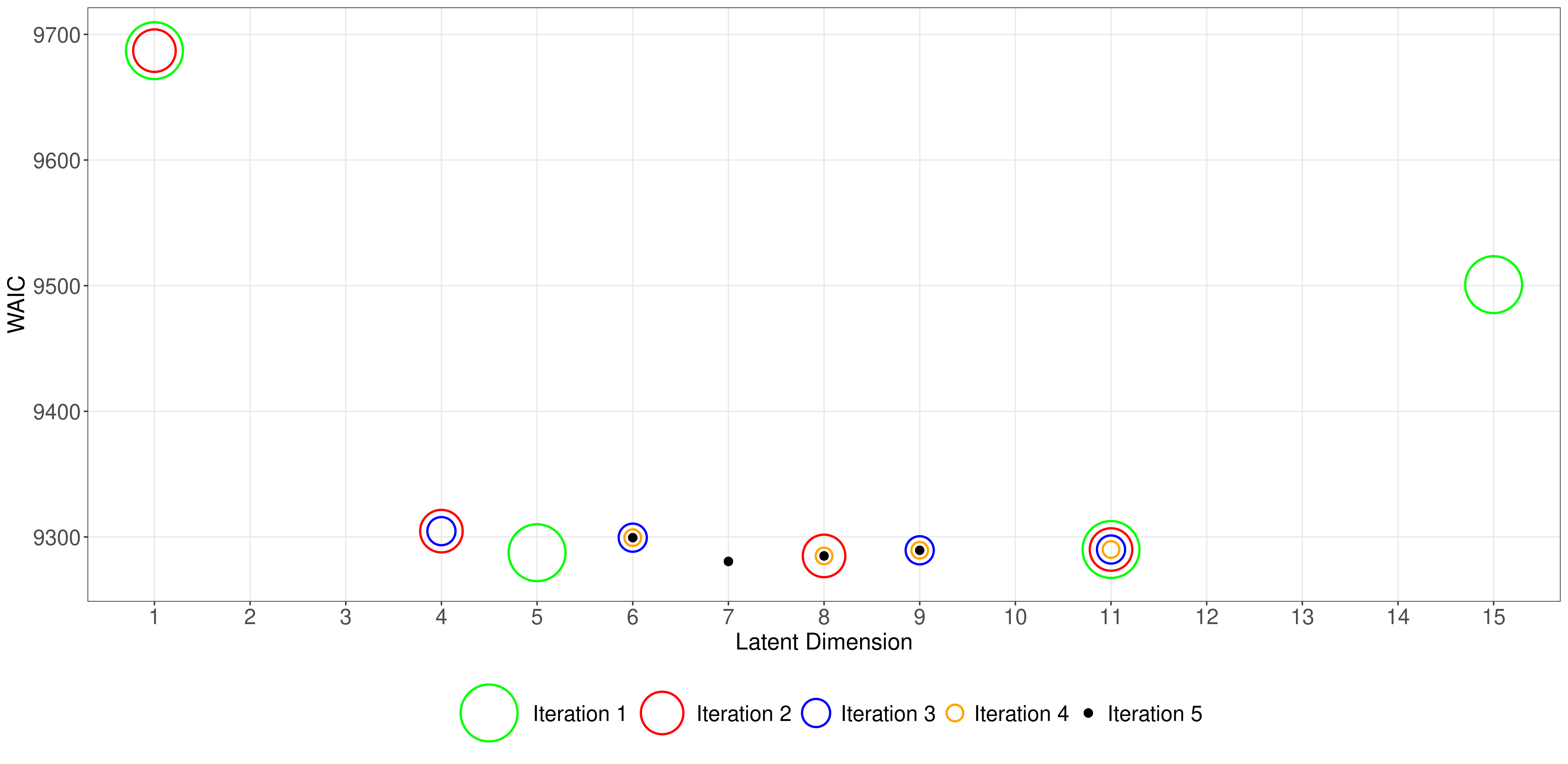}
    \caption{Iterations of the ternary search to minimize WAIC.}
    \label{fig:waic_ternary}
\end{figure}

To select a latent dimension and avoid fitting the model for every valid dimension, $u=1,\ldots,23$, we performed a ternary search between a minimum and maximum considered dimension, $u_\text{min}=1$, $u_\text{max}=15$, attempting to minimize WAIC (Figure \ref{fig:waic_ternary}). The WAIC surface is noisy and not strictly convex, so a ternary search is not guaranteed to find a global minimum, but can find a reasonable model while considering fewer candidate models. Among the nine candidate models, $u = 7$ produced the lowest WAIC.

Sampling spatial weights $\bw_j;\,j=1,\ldots, u$ dominated the computation time, so the total computation time was almost exactly linear with dimension $u$. On average, a single Gibbs cycle across all unknowns consumed 0.12 seconds for the $u=1$ model, 0.59 seconds for the $u=7$ model, and 1.48 seconds for the $u = 15$ model, processing on a AMD Ryzen Threadripper PRO 7985WX.

\begin{figure}
    \centering
    \begin{subfigure}[b]{\textwidth}
        \centering
        \includegraphics[width = 0.8 \textwidth]{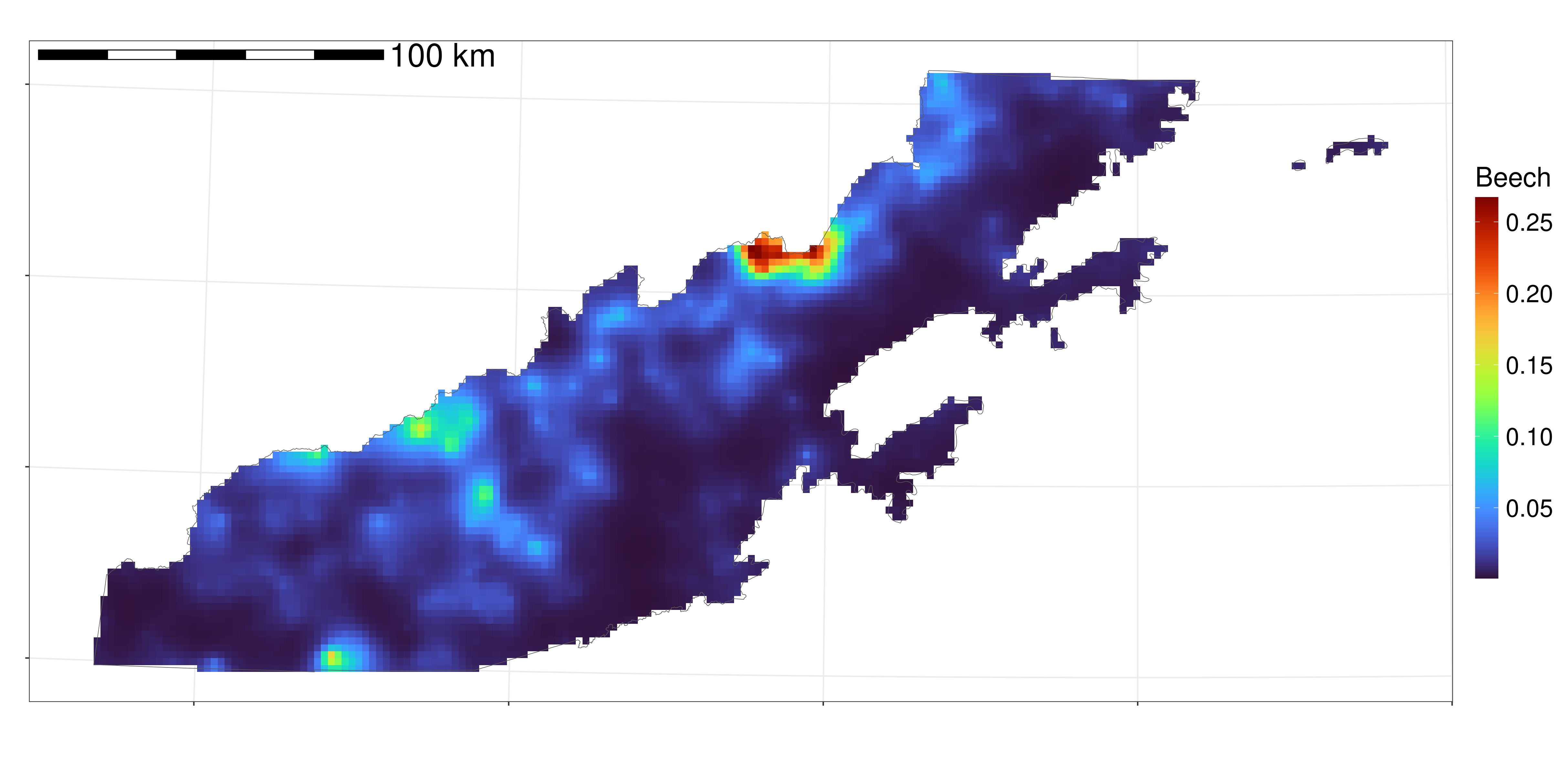}
        \caption{Probability of Beech dominance at the observation level.}
        \label{fig:preda}
    \end{subfigure} \\
    \begin{subfigure}[b]{\textwidth}
        \centering
        \includegraphics[width = 0.8 \textwidth]{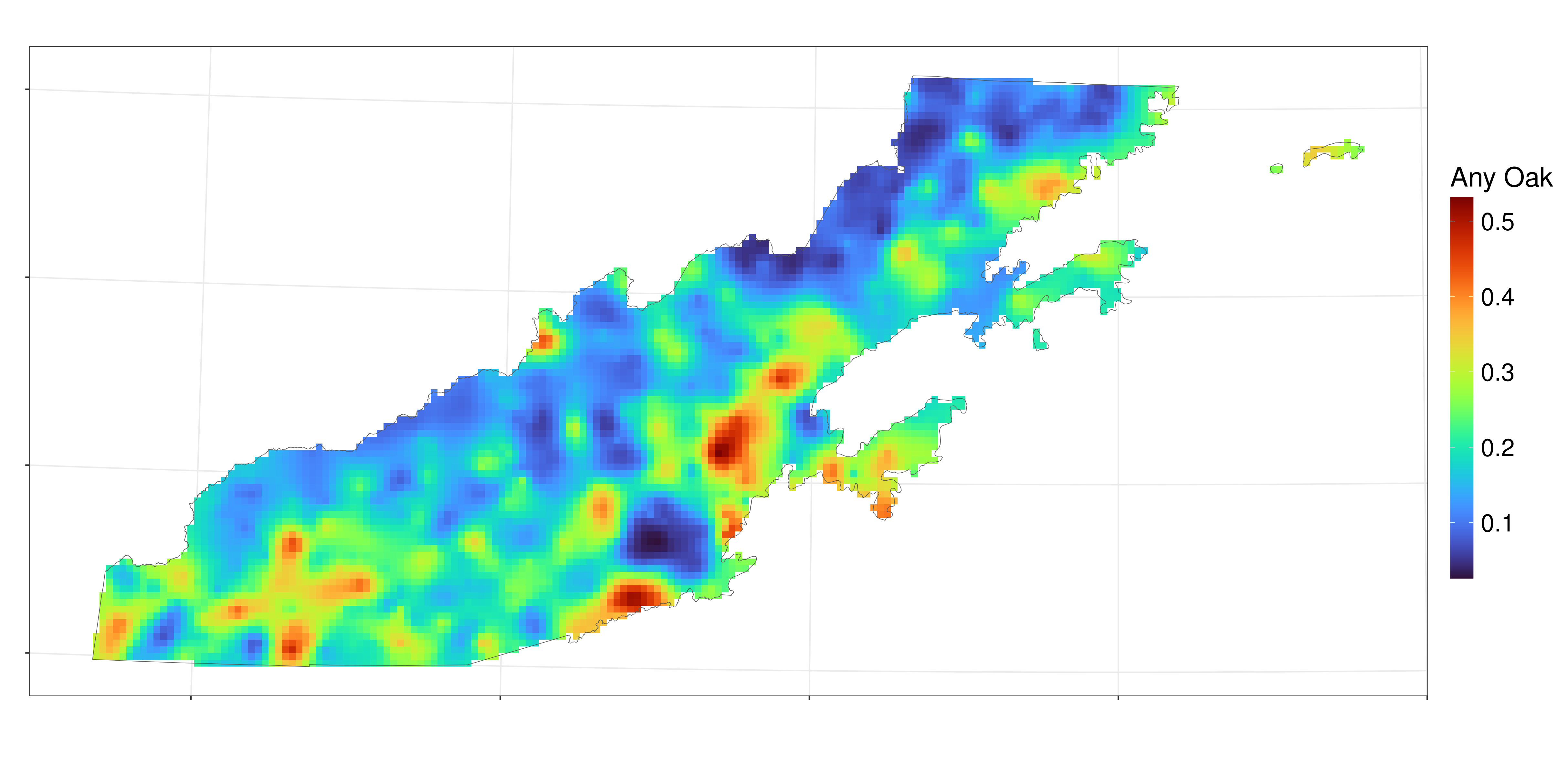}
        \caption{Probability of dominance from any Oak class (4 classes) at the observation level}
        \label{fig:predb}
    \end{subfigure} \\
    \begin{subfigure}[b]{\textwidth}
        \centering
        \includegraphics[width = 0.8 \textwidth]{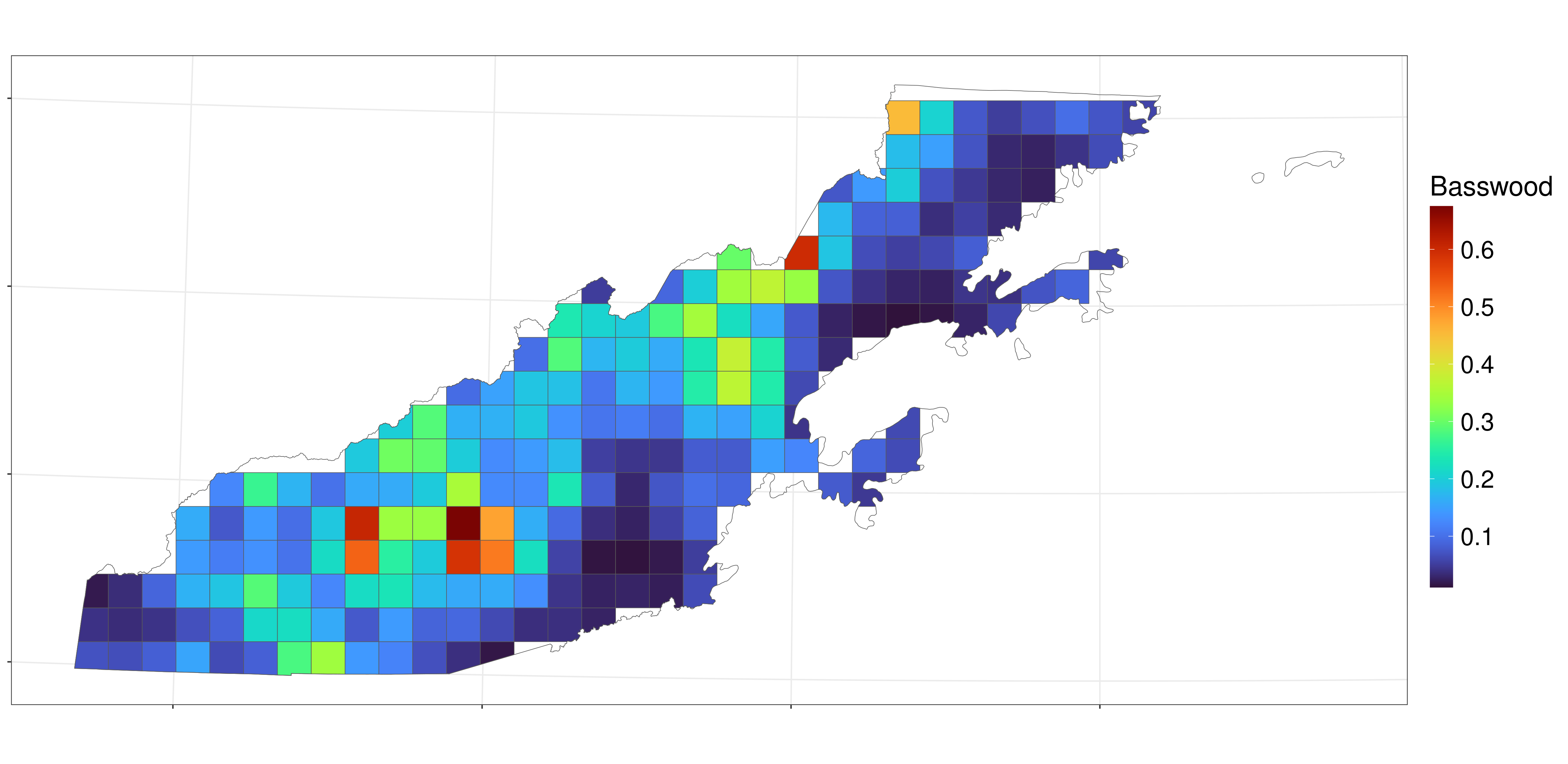}
        \caption{Probability of any Basswood dominance within $10 \times10$ km areas.}
        \label{fig:predc}
    \end{subfigure} \\
    \caption{Example posterior predictions across the study area, demonstrating the ability to predict probabilities for single classes, unions of classes, and area summaries.}
    \label{fig:placeholder}
\end{figure}

Using the $u=7$ model, we generated predictive samples of $\bp(\bs)$ and $\by(\bs)$ on a regular 1 km grid. These samples can be used flexibly to characterize our posterior knowledge of class occurrences across the study domain, for instance, to predict probabilities for individual classes (Fig.~\ref{fig:preda}), probabilities for a union of classes (Fig.~\ref{fig:predb}), and probabilities of any class occurrence within an area (Fig.~\ref{fig:predc}).

%%%%%%%%%%%%%%%%%%%%%%%%%%%%%%%
\section{Discussion} \label{sec:disc}
% A summary of the model and motivation
We presented a spatial multinomial model where the spatial effects are expressed as a reduced set of linear effects, expanding the popular factor model to the multinomial setting. Because the reduced rank precludes common Bayesian computational techniques for multinomial models, we proposed a Gibbs sampler based on Laplace approximation proposal distributions. The model and inference was demonstrated for categorical data through simulation and a real data analysis.

% dimension selection
The selection of the latent dimension $u$ is a primary challenge to the proposed model. While our simulations show that common model selection techniques are effective at selecting an efficient model when a reduced-rank structure is present, this involves fitting the model for all considered dimensions. Techniques like a ternary search between a minimum and maximum considered dimension, as used in Section \ref{sec:rda}, can reduce the pool of candidate models, but the procedure remains time consuming for data with many classes. The development of effective heuristics that avoid repeatedly fitting the model for varying dimensions would be a useful research avenue.  

% Laplace approximations
Within the Gibbs sampler, the latent spatial effects are sampled through a Laplace approximation proposal to a Metropolis-Hastings step. In our experience, the key driver to the accuracy of the proposal and the Metropolis-Hastings acceptance rates is the marginal precision of the spatial effects. Small precisions induce skewed conditional posteriors and biased, inaccurate proposal distributions. If the data likelihood favors small precisions and the computational cost per sample is high (large Gaussian process rank $k$ and/or many classes) the proposed sampler may not be practical. \cite{rue2009approximate} (Section 3.2.3) proposes a third-order Taylor expansion of the log-density to motivate a multivariate skew-normal approximation, correcting some of the bias present in the Laplace approximation (a second-order expansion) and accounting for skew. Using a skew-normal distribution to generate proposals could improve acceptance rates and make the sampler feasible for smaller precisions. Using either the second or third order expansion, the computational cost per sample is dominated by the Cholesky decomposition of a posterior precision matrix of the form $\omega\bQ+\bB^T\bD\bB$, where $\omega\bQ$ is the prior spatial precision matrix, $\bB$ is the design matrix of basis functions, and $\bD$ is a diagonal matrix. The predictive process \citep{banerjee2008gaussian} produces dense prior precision and basis matrices, but other Gaussian process models produce sparse matrices (see Section 2.3 of \cite{heaton2019case} for a review of such models), allowing sparse decompositions and faster inference.

% Multinomial
The model was only demonstrated for categorical data, but is applicable to multinomial models where either the number of trials is known across the spatial domain or if only class probabilities are of interest and not absolute counts. The sampler in Appendix \ref{app:gibbs} is general to multinomial models with arbitrary trials and shows the approximate posterior precision of the latent effects to increase linearly with the number of observed trials. If absolute counts are of interest and the number of trials unknown at unobserved locations, the likelihood is better specified as $J$ conditionally independent Poisson distributions. In this case, using a logarithm link function but the same factored linear predictor as (\ref{eq:redranklogits}), the gradients and Hessians of the conditional log-densities are similar to the multinomial case and a similar Gibbs sampler could be applied.

\section*{Code and data}
All algorithms used in this study are available as a \texttt{Julia} package at \url{https://github.com/PaulBMay/SpatialMultinomial.jl}. Forest Inventory and Analysis data is publicly available through the FIA DataMart (\url{https://research.fs.usda.gov/products/dataandtools/tools/fia-datamart}) or through \texttt{R} package `FIESTA' \citep{Frescino_2023}.

\section*{Funding}
This research was partially supported by the National Science Foundation and the National Geospatial-Intelligence Agency (\#2428037) and the USDA Forest Service (\#FO-SFG-2673, \#24-JV-11242305-111). The opinions, findings, and conclusions in this publication are those of the authors and do not represent any official policy or determination by the National Science Foundation, National Geospatial-Intelligence Agency, or US Department of Agriculture.

\bibliographystyle{apalike}
\bibliography{bibliography}

\appendix

\section{Gibbs Sampler} \label{app:gibbs}

Given $n$ observations at locations $\bs_1,\ldots,\bs_n\in\mathcal{D}$, let $\bY=[\by_1\cdots\by_n]^T$ and $\bP=[\bp_1\cdots\bp_n]^T$ be $n\times(J-1)$ matrices of multinomial observations, excluding the control class. Let $N_i\,;\,i=1,\ldots,n$ represent the number of multinomial trials for each observation and $\boldsymbol{N}=\diag(N_1,\ldots,N_n)$. Let $\bB(\phi) = [\bb_1(\phi) \cdots \bb_n(\phi)]^T$ be the $n\times k$ matrix of basis functions. Finally, let $\bgamma$ be the vector of unconstrained entries in $\bGamma$, concatenated column-wise. The full hierarchical data model is
\begin{align}
    \bY | \bP, \boldsymbol{N} &\iid \mathrm{Multinomial}(\bP,\boldsymbol{N}), \\
    \bP | \bPsi &= \mathrm{softmax_J(\bPsi)}, \\
    \bPsi|\bmu,\bW,\bgamma, \phi &= \bone\bmu^T+\bB(\phi)\bW\bGamma(\bgamma)^T,\\
    ~~ \notag \\
    \bw_j|\omega_j,\phi &\sim \mathrm{MVN}(\bzero, \omega_j^{-1}\bQ(\phi)^{-1})\quad\text{for}~j=1,\ldots,u, \\
    \omega_j&\sim\mathrm{Gamma}(\alpha_{\omega, i}, \beta_{\omega,i})\quad\text{for}~j=1,\ldots,u, \\
    \phi &\sim \mathrm{Gamma}(\alpha_\phi, \beta_\phi), \\
    \bmu &\sim \mathrm{MVN}(\bm_\mu, \bQ_\mu), \\
    \bgamma &\sim \mathrm{MVN}(\bm_\gamma, \bQ_\gamma).
\end{align}

The Laplace approximations for the conditional posteriors of $\bw_j, \bgamma, \bmu$ require the gradient, $\bd(\cdot)$, and Hessian, $\bH(\cdot)$, of the relevant log-density. Let $\boldsymbol{E}_\gamma$ be the selection matrix such that $\bgamma = \boldsymbol{E}_\gamma\vecc(\bGamma)$. Temporarily omitting all function notation for brevity, e.g. $\bB\equiv \bB(\phi)$, we have 
\begin{align}
    \bd(\bw_j) &= \bB^T\left(\bY-\boldsymbol{N}\bP\right)\bGamma_j-\omega_j\bQ\bw_j,\\
    \bH(\bw_j) &= -\omega_j\bQ-\bB^T\mathrm{diag}\{\boldsymbol{c}\}\bB,~~\text{where}~~ c_i = N_i\left[\sum_{\ell=1}^{J-1}\left\{p_{i\ell}\Gamma_{\ell j}^2\right\}-\left(\sum_{\ell=1}^{J-1}p_{i\ell}\Gamma_{\ell j} \right)^2\right], \\
    \bd(\bgamma) &= \boldsymbol{E}_\gamma\vecc(\bW^T\bB^T(\bY-\boldsymbol{N}\bP))-\bQ_\gamma(\bgamma -\bm_\gamma), \\
    \bH(\bgamma) &= -\bQ_\gamma - \boldsymbol{E}_{\bgamma} \left(\sum_{i=1}^n N_i\left[(\diag(\bp_i)-\bp_i\bp_i^T) \otimes \boldeta_i\boldeta_i^T\right]\right)\boldsymbol{E}_\gamma^T,~~\text{where}~~\boldeta_i=\bW\bb_i, \\
    \bd(\bmu) &= (\bY-\boldsymbol{N}\bP)^T\bone - \bQ_\mu(\bmu - \bm_\mu), \\
    \bH(\bmu) &= -\bQ_\mu - \sum_{i=1}^n N_i\left[\diag(\bp_i)-\bp_i\bp^T \right].
\end{align}
To generate a proposal from the Laplace approximation, we find the mode of the conditional posterior using Newton-Raphson. Using $\bgamma$ for exposition, the mode $\hat{\bgamma}$ is found by iterating
\begin{equation}
    \bgamma^{(\ell+1)} = \bgamma^{(\ell)} - \bH\left(\bgamma^{(\ell)}\right)^{-1}\bd\left(\bgamma^{(\ell)}\right),
\end{equation}
until some stopping criterion is achieved. A proposal is then generated,
\begin{equation}
    \bgamma^* \sim \mathrm{MVN}\left(\hat{\bgamma}, -\bH(\hat{\bgamma})^{-1}\right),
\end{equation}
and either accepted or rejected in a Metropolis-Hastings step. A cycle of the Gibbs sampler then iterates through the following:
\begin{itemize}
    \item Sample $\bw_j|\cdots$ via a Laplace proposal to a Metropolis-Hastings step for $j=1,\ldots,u$.
    \item Sample $\bgamma | \cdots$ via a Laplace proposal to a Metropolis-Hastings step.
    \item Sample $\bmu | \cdots$ via a Laplace proposal to a Metropolis-Hastings step.
    \item Sample $\omega_j |\cdots \sim \mathrm{Gamma}(\tilde{\alpha}_{\omega,j}, \tilde{\beta}_{\omega,j})$ with
    \begin{align}
        \tilde{\alpha}_{\omega,j} &= \alpha_{\omega,j} + \frac{n}{2} \\
        \tilde{\beta}_{\omega,j} &= \beta_{\omega,j} + \frac{\bw_j^T\bQ(\phi)\bw_j}{2}
    \end{align}
    for $j=1,\ldots,u$.
    \item Sample $\phi|\cdots$ via a Metropolis-Hastings step.
\end{itemize}

\end{document}